\documentclass[%
 reprint,
 amsmath,amssymb,
 aps,
]{revtex4-1}
\usepackage{physics}
\usepackage{graphicx}
\usepackage{dcolumn}
\usepackage{bm}

\begin{document}

\preprint{APS/123-QED}

\title{Ultrafast Energy Transfer in the Metal Nanoparticles-Graphene Nanodisks-Quantum Dots Hybrid Systems
}

\author{Mariam Tohari}
\author{Andreas Lyras}%
\author{Mohamad AlSalhi}%

\affiliation{%
 Department of Physics and Astronomy, College of Science, King Saud University, Riyadh, P. O. Box 11451, Saudi Arabia\\Research Chair on Laser Diagnosis of Cancers, College of Science, King Saud University, Saudi Arabia
}%

\date{\today}

\begin{abstract}
Hybrid nanocomposites can offer a wide range of opportunities to control the light-matter interaction and electromagnetic energy flow at the nanoscale, leading to exotic optoelectronic devices. We study theoretically the dipole-dipole interaction in noble metal nanoparticles-graphene nanodisks-quantum dots hybrid systems in the optical region of the electromagnetic spectrum. The quantum dot is assumed to be a three-level atom interacting with ultrashort control and probe pulses in a $\Lambda $ configuration. The dynamics of the system are studied by numerically solving for the time evolution of the density matrix elements. We investigate the rate of energy exchange between surface plasmon resonances of the graphene nanodisks and excitons of the quantum dots in the presence of metal nanoparticles at steady state and for specific geometrical conditions of the system. Ultrafast population dynamics are obtained with a large energy exchange rate significantly depending on the size of metal nanoparticles. The power transfer can be controlled by varying the center-to-center distances between the components of the system, and their positions with respect to each other. We also find that the rate of energy transfer within the system is governed by the probe field Rabi frequency, enhanced by the dipole-dipole interaction. 
\end{abstract}

\pacs{Valid PACS appear here}
\maketitle


\section{Introduction}
\label{intro}

Plasmonics are widely studied due to their applications in ultrasensitive optical biosensing \cite{anker2008biosensing,rodrigo2015mid}, photonics metamaterial \cite{zheludev2011roadmap,boltasseva2011low}, light harvesting \cite{atwater2010plasmonics}, optical nanoantennas \cite{ni2012broadband}, photocatalysis \cite{awazu2008plasmonic,zhang2013plasmonic} and quantum information \cite{gonzalez2011entanglement,tame2013quantum}. Specifically, plasmonic nanostructures have the ability to confine light within sub-wavelength dimensions, potentially contributing to significant enhancement of optoelectronics devices quality measures \cite{bao2012graphene,jarrahi2015advanced,li2016controlling}.

Localized surface plasmons of metal nanoparticles (MNPs) provide a mechanism for resonance excitations based on the ability of MNPs to scatter and absorb light \cite{hutter2004exploitation,chon2013nanoplasmonics}. The extinction cross section and the spectral position of resonances are controlled by the size and shape of MNPs as well as the dielectric environment \cite{noguez2007surface}. Although, noble metals are considered as the best plasmonic materials that guide visible light at the nanoscale \cite{west2010searching}, they are hardly tunable \cite{sreeprasad2013noble} and suffer large ohmic losses \cite{khurgin2012reflecting}. On the other hand, graphene plasmons exhibit tighter confinement and relatively long propagation distances due to the high mobility of their charge carriers \cite{bolotin2008ultrahigh,gonccalves2016introduction}. Moreover, surface plasmons of graphene can be tunable via electrostatic gating \cite{guo2011graphene} which carries potential for technological applications. Graphene nanodisks (GNDs) are suitable to boost light-matter interactions at higher photon energies due to their ability to confine light in all dimensions. Moreover, the long life time of their plasmons enables them to play the role of effective resonators \cite{ezawa2008coulomb,minovkoppensich2011graphene}. Because of the nonlinearity of their plasmons, plasmon blockade effects emerge naturally in the optical response of GND leading to nonlinear optical absorption cross sections \cite{ezawa2008coulomb,manjavacas2012plasmon}.

There is considerable interest in developing nanoscale optoelectronic devices by combining nanomaterials into hybrid structures. The enhancement of light-matter interaction in metal-graphene hybrid systems has been studied by several research groups in order to reduce the losses \cite{rast2013stratified}, investigate surface Raman scattering \cite{schedin2010surface}  and control the nanoscale graphene plasmonic circuits \cite{alonso2014controlling}. The plasmonic structures for both noble metals and graphene have the ability to enhance the optical properties of quantum emitters in their vicinity, such as quantum dots (QDs) and molecules, due to  plasmon-induced field enhancement \cite{heliotis2006hybrid,govorov2006exciton,govorov2007theory,komarala2008surface,dombi2009surface,kanemitsu2011energy,cox2012dipole,biehs2013large,carreno2014plasmon}. Thus, a strong coupling is induced between the optical excitations of quantum emitters and plasmonic materials, i.e. excitons and plasmons respectively, when they are resonant causing energy transfer via dipole-dipole interaction (DDI) \cite{govorov2007theory}. The energy transfer between excitons and plasmons in semiconductor-metal and semiconductor-graphene hybrid systems has been studied both theoretically and experimentally in different coupling regimes \cite{heliotis2006hybrid,govorov2006exciton,govorov2007theory,komarala2008surface,dombi2009surface,kanemitsu2011energy,cox2012dipole,biehs2013large,carreno2014plasmon}. A femtowatt energy exchange rate has been demonstrated between GND and QD via DDI in the infrared regime of the electromagnetic spectrum \cite{cox2012dipole}. It was found that the graphene plasmonics can enhance the energy transfer in the infrared \cite{biehs2013large}. However, localized surface plasmon resonances (LSPRs) in GND can be controlled via the size and doping level to reach the optical region which could be important for certain applications. Thanks to recent developments in  2D materials doping techniques, extremely high doping levels can be achieved ($\succ10^{14} cm^{-2}$) \cite{paradisi2015space}.

\begin{figure}
\includegraphics{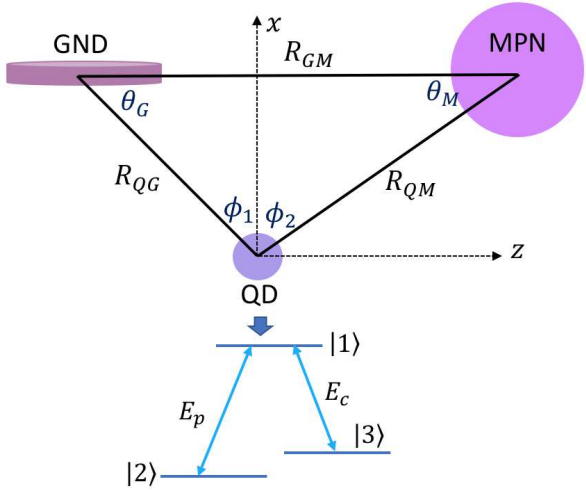}
\caption{\label{fig:1} The schematic illustration of the proposed model of MNP-GND-QD hybrid system.}
\end{figure}

Motivated by the experimental results, that the plasmons of graphene can be launched and controlled with resonant metal antennas \cite{alonso2014controlling}, we will investigate the energy transfer in the optical regime within the MNP-GND-QD hybrid system schematically depicted in Fig.~\ref{fig:1}. The DDI between the components of the system and the related enhancement in the interaction will be studied, as well as the corresponding dynamics by numerically solving for the time evolution of the density matrix elements. The self-assembled QD is modeled as a three- level atom of $\Lambda $ configuration in which two excitonic transitions are induced via probe and control ultrashort Gaussian pulses \cite{chu2007three,hohenester2000coherent}. We will investigate the rate of energy exchange at steady state under various conditions related to the structure of the system. 

The paper is organized as follows; in Sec.~\ref{sec2} the model is established, i.e., the dipole-dipole Hamiltonian of the system, assuming the rotating wave approximation, and the corresponding time evolution of density matrix elements that describes the dynamics of the system. Sec.~\ref{sec3} displays the numerical simulations and discusses the dynamics and energy transfer rate under different conditions related to the specific system and Sec.~\ref{sec4} summarizes the main conclusions.

\section{Theoretical Formalism}
\label{sec2}
We investigate the dynamics and the rate of energy exchange within the proposed MNP-GND-QD hybrid system. We consider a spherical MNP of dielectric constant $\epsilon_{M}$ and shape-dependent polarizability $\alpha_{M}$ embedded in dielectric medium of $\epsilon_{b}$ at center-to-center distance $R_{GM}$ from the GND of $\epsilon_{G}$ and $\alpha_{G}$ correspondingly. A self-assembled QD, modeled as a three-level $\Lambda $ type atom \cite{chu2007three}, is situated at center-to-center distances $R_{QM}$ and $R_{QG}$ from the MNP and GND respectively. Angles $\theta_{M}$, $\theta_{G}$, $\phi_{1}$ and $\phi_{2}$ , as illustrated in Fig.~\ref{fig:1}, define the geometry of the set-up. By considering that $\mu_{12}$ ($\mu_{13}$) lies along the x direction (z direction), the applied x-polarized probe field (z-polarized control field) will induce the excitonic transition $\ket{1}\leftrightarrow\ket{2}$ ($\ket{1}\leftrightarrow\ket{3}$).The probe and control fields create excitons within the QD and localized surface plasmons in both the MNP and GND. When the energy of the surface plasmons is resonant with that of the excitons, then a strong coupling between excitons and plasmons can occur \cite{singh2016excitonic,torma2014strong,baieva2012strong,wang2016dynamics}.  The DDI between the components of the system can be used to study the coupling between their optical excitations.

The DDI of QD induced by probe field (control field) applied along x-direction (z-direction) is described for our system by the total dipole field:
\begin{eqnarray}
E_{DDI}^{j}=\frac{\hbar}{\mu_{1i}}\left[\Omega_{j}\left(\Pi_{x,z}+\Phi_{x,z}\right)+\Lambda_{x,z}\rho_{1i}\right]
\label{eq:1}.
\end{eqnarray}
where $j$ and $i$ indices indicate the interacting field,$j=p (c)$ for probe field (control field), and the corresponding atomic level $\left(i=2,3\right)$ coupled to excited state 1 through the atomic coherence $ \rho_{1i}$ respectively. The first term represents the dipole fields from GND and MNP induced by field j while the second term arises when the field j polarizes the QD which in turn polarizes GND and MNP \cite{singh2016excitonic}. In Eq.~(\ref{eq:1}), $\Omega_{j}=E_{j}. \mu_{1i}/\hbar$ is the Rabi frequency of the field j. For small size of the components, much smaller than the wavelength of the incident light, $ \Pi_{x,z}$,$\Phi_{x,z}$, and $\Lambda_{x,z}$ are given based on the near-field approximation \cite{novotny2006principles} by:
\begin{widetext}
\begin{subequations}
\label{eq:2}
\begin{eqnarray}
\Pi_{x}=\frac{1}{4\pi \epsilon^{*}}\left[\frac{\alpha_{G}^{x}\left(3cos \phi_{1}-1\right)}{R_{QG}^3}+\frac{\alpha_{M}\left(3cos \phi_{2}-1\right)}{R_{QM}^3}\right],\label{2a}
\end{eqnarray}
\begin{equation}
\Lambda_{x}=\frac{\mu_{12}^{2}}{\left(4\pi \epsilon^{*}\right)^{2}\hbar \epsilon_{0}\epsilon_{b}}\left[\frac{\alpha_{G}^{x}\left(3cos \phi_{1}-1\right)^2}{R_{QG}^6}+\frac{\alpha_{M}\left(3cos \phi_{2}-1\right)^2}{R_{QM}^6}\right],\label{2b}
\end{equation}
\begin{equation}
\Phi_{x}=\frac{-\alpha_{G}^{x}\alpha_{M}}{\left(4\pi \epsilon^{*}\right)^{2} R_{GM}^3 }\left[\frac{3cos \phi_{1}-1}{R_{QG}^3}+\frac{3cos \phi_{2}-1}{R_{QM}^3}\right],\label{2c}
\end{equation}
\begin{equation}
\Pi_{z}=\frac{1}{4\pi \epsilon^{*}}\left[\frac{\alpha_{G}^{z}\left(3 cos\theta_{G} -1\right)}{R_{QG}^3}+\frac{\alpha_{M}\left(3 cos\theta_{M} -1\right)}{R_{QM}^3}\right],\label{2d}
\end{equation}
\begin{equation}
\Lambda_{z}=\frac{\mu_{13}^{2}}{\left(4\pi \epsilon^{*}\right)^{2}\hbar \epsilon_{0}\epsilon_{b}}\left[\frac{\alpha_{G}^{z}\left(3cos \theta_{G}-1\right)^2}{R_{QG}^6}+\frac{\alpha_{M}\left(3cos \theta_{M}-1\right)^2}{R_{QM}^6}\right],\label{2e}
\end{equation}
\begin{equation}
\Phi_{z}=\frac{2\alpha_{G}^{z}\alpha_{M}}{\left(4\pi \epsilon^{*}\right)^{2} R_{GM}^3 }\left[\frac{3cos \theta_{G}-1}{R_{QG}^3}+\frac{3cos \theta_{M}-1}{R_{QM}^3}\right],\label{2f}
\end{equation}
\end{subequations}
\end{widetext}

In the above expressions, $\mu_{1i}$ is the dipole moment of excitonic transition $\ket{1}\leftrightarrow\ket{i}$ and $\epsilon^{*} =\frac{\left(2\epsilon_{b}+\epsilon_{q}\right)}{3\epsilon_{b}}$ is the effective dielectric constant where $\epsilon_{b}$ and $\epsilon_{q}$ are the dielectric constants of background and QD respectively. $\alpha_{G}^{x,z}$represents the shape-dependent polarizability of GND along x and z directions given by \cite{singh2016excitonic}:
\begin{equation}
\alpha_{G}^{x,z}=\frac{4\pi V_{G}\left[\epsilon_{G}(\omega )-\epsilon_{b}\right]}{3\epsilon_{b}+3 \zeta_{x,z}\left[\epsilon_{G}(\omega)-\epsilon_{b}\right]}
\label{eq:3}
\end{equation}
 where $V_{G}$ is the volume of GND and $\zeta_{x,z}$ is the depolarization factor of GND given for $L_{y}=L_{z}\gg L_{x}$ by \cite{sarid2010modern}:
 \begin{equation}
\zeta_{x}=1-\frac{\pi L_{x}}{2L_{z}} \quad \textrm{and} \quad\zeta_{y}=\zeta_{z}=\frac{\pi L_{x}}{4L_{z}}
\label{eq:4}
\end{equation}
The dielectric function of graphene includes contributions from both inter-band and intra-band transitions. The former is temperature- dependent (T) and significant for optical wavelengths as noted from the following expression for the dielectric function of graphene \cite{premaratne2017theory}: 
\begin{eqnarray}
{\epsilon_{G}(\omega)}=&&1+\frac{i e^{2}}{8 \epsilon_{0} \hbar d \omega}\nonumber\\
&&\left[\tanh \left(\frac{\hbar \omega+2E_{F}}{4k_{B}T}\right) +\tanh \left(\frac{\hbar \omega-2E_{F}}{4k_{B}T}\right)\right]\nonumber\\
&& + \frac{e^{2}}{8 \pi \hbar\epsilon_{0} d\omega} \ln \left[\frac{\left(\hbar \omega +2E_{F}\right)^2}{\left(\hbar \omega -2E_{F}\right)^2+\left(2k_{B}T\right)^{2}}\right] \nonumber\\
&& -\frac{e^{2}}{ \pi \hbar \epsilon_{0} d \omega}\left(\frac{E_{F}}{\hbar \omega+i \hbar \gamma}\right)
\label{eq:5}
\end{eqnarray}
where  $E_{F}$ is the Fermi energy, $\gamma$ is intra-band scattering rate and $k_{B}$ is the Boltzmann constant. 

The shape-dependent polarizability of the MNP is given in terms of its volume $V_{M}$ and dielectric function $\epsilon_{M}(\omega)$ for metal of plasma frequency $\omega_{p}$ and damping $\gamma_{M}$ by \cite{chon2013nanoplasmonics}:
 \begin{equation}
\alpha_{M}=V_{M}\left[\frac{\epsilon_{M}(\omega)-\epsilon_{b}}{\epsilon_{M}(\omega)+2\epsilon_{b}}\right]
\label{eq:6}
\end{equation}
where $\epsilon_{M}(\omega)=\epsilon_{\infty}-\frac{\omega_{p}^2}{\omega^2+i\omega\gamma_{M}}$. The DDI Hamiltonian term corresponding to $E_{DDI}^{j}$ can be written as:
 \begin{equation}
H_{DDI}=-\sum_{i,j} \hbar \left[\Omega_{j}\left(\Pi_{x,z}+\Phi_{x,z}\right)+\Lambda_{x,z}\rho_{1i}\right]\sigma_{1i}+H.C.
\label{eq:7}
\end{equation}
The total Hamiltonian of this system is given in the rotating wave approximation frame in terms of one and two-photon detuning as \cite{scully1997quantum}:
\begin{eqnarray}
{H^{RWA}}=&&\hbar(\Delta_{p}\sigma_{11}+\Delta_{2}\sigma_{33})\nonumber\\
&&-\hbar\left[\Omega_{p}(\Pi_{x}+\Phi_{x})+\Lambda_{x}\rho_{12}\right]\sigma_{12}\nonumber\\
&&- \hbar\left[\Omega_{c}(\Pi_{z}+\Phi_{z})+\Lambda_{z}\rho_{13}\right]\sigma_{13}+H.C.
\label{eq:8}
\end{eqnarray}
where $\Delta_{2}=\Delta_{p}-\Delta_{c}$ is the two-photon detuning. $\Delta_{p}=\omega_{12}-\omega_{p}$ and $\Delta_{c}=\omega_{13}-\omega_{c}$ are the detuning of probe and control fields respectively. $\sigma_{11}$ and $\sigma_{33}$ represent the projection operators onto the lower and upper levels whereas $\sigma_{1i}$ give the flip operators connected to the optical transitions. Now, we can use the density matrix method to investigate the energy transfer between QD and GND in the presence of MNP with Liouvillian  given for this system by \cite{scully1997quantum}:
\begin{eqnarray}
{L_{\rho}}=&&\frac{\gamma_{13}}{2}(\rho \sigma_{11}+\sigma_{11} \rho-2\sigma_{31}\rho\sigma_{13})\nonumber\\
&& +\frac{\gamma_{12}}{2}(\rho \sigma_{11}+\sigma_{11} \rho-2\sigma_{21}\rho\sigma_{12})\nonumber\\
&& +\frac{\gamma_{32}}{2}(\rho \sigma_{33}+\sigma_{33} \rho-2\sigma_{23}\rho\sigma_{32})
\label{eq:9}
\end{eqnarray}
$\gamma_{1i}$ stand for the spontaneous decay rates of QD and $\gamma_{32}$ accounts for lower states' dephasing. The time evolution of the density matrix elements now reads:
\begin{widetext}
\begin{subequations}
\label{eq:10}
\begin{eqnarray}
\dot{\rho}_{13}=-\left[\left(\frac{\gamma_{13}}{2}+\frac{\gamma_{12}}{2}\right)+i\left(\Delta_{c}-\Lambda_{z}\left(\rho_{33}-\rho_{11}\right)\right)\right]\rho_{13}+i\Omega_{c}\left(\Pi_{z}+\Phi_{z}\right)\left(\rho_{33}-\rho_{11}\right)+i\left[\Omega_{p}\left(\Pi_{x}+\Phi_{x}\right)+\Lambda_{x}\rho_{12}\right]\rho_{23},\label{1a}
\end{eqnarray}
\begin{equation}
\dot{\rho}_{12}=-\left[\left(\frac{\gamma_{13}}{2}+\frac{\gamma_{12}}{2}\right)+i\left(\Delta_{p}-\Lambda_{x}\left(\rho_{22}-\rho_{11}\right)\right)\right]\rho_{12}+i\Omega_{p}\left(\Pi_{x}+\Phi_{x}\right)\left(\rho_{22}-\rho_{11}\right)+i\left[\Omega_{c}\left(\Pi_{z}+\Phi_{z}\right)+\Lambda_{z}\rho_{13}\right]\rho_{32},\label{1b}
\end{equation}
\begin{equation}
\dot{\rho}_{32}=-\left(\frac{\gamma_{32}}{2}+i\Delta_{2}\right)\rho_{32}+i\left[\Omega_{c}^{*}\left(\Pi_{z}^{*}+\Phi_{z}^{*}\right)+\Lambda_{z}^{*}\rho_{31}\right]\rho_{12}-i\left[\Omega_{p}\left(\Pi_{x}+\Phi_{x}\right)+\Lambda_{x}\rho_{12}\right]\rho_{31},\label{1c}
\end{equation}
\begin{equation}
\dot{\rho}_{11}=-\left(\gamma_{12}+\gamma_{13}\right)\rho_{11}+i\left[\Omega_{c}\left(\Pi_{z}+\Phi_{z}\right)+\Lambda_{z}\rho_{13}\right]\rho_{31}+i\left[\Omega_{p}\left(\Pi_{x}+\Phi_{x}\right)+\Lambda_{x}\rho_{12}\right]\rho_{21}+c.c.,\label{1d}
\end{equation}
\begin{equation}
\dot{\rho}_{22}=\gamma_{12}\rho_{11}+\gamma_{32}\rho_{33}-i\left[\Omega_{p}\left(\Pi_{x}+\Phi_{x}\right)+\Lambda_{x}\rho_{12}\right]\rho_{21}+c.c.,\label{1e}
\end{equation}
\begin{equation}
\dot{\rho}_{33}=\gamma_{13}\rho_{11}-\gamma_{32}\rho_{33}-i\left[\Omega_{c}\left(\Pi_{z}+\Phi_{z}\right)+\Lambda_{z}\rho_{13}\right]\rho_{31}+c.c.,\label{1f}
\end{equation}
\end{subequations}
\end{widetext}
It is remarkable from the first two equations of  Eqs.~(\ref{eq:10}) that $Re\left[\Lambda_{x,z}(\rho_{ii}-\rho{11})\right]$ represents the non-radiative energy shift due to the dipole term $\Lambda_{x,z}$ while $Im\left[\Lambda_{x,z}(\rho_{ii}-\rho{11})\right]$ gives the decay rate due to the dipole term. This means that the spontaneous emission of the QD is enhanced due to the hybrid plasmonic cavity by $\Lambda_{x,z}$. Thus, we can control the spontaneous emission via the geometrical and structural parameters of the system. Moreover, the strength of the interaction between excitons and plasmons, described by the Rabi frequency, is enhanced by a factor$\left(\Pi_{x}+\Phi_{x}\right)$ due to the DDI leading to strong coupling between excitons and plasmons for $|\Pi_{x}+\Phi_{x}|>1$.
\begin{figure}[b]
\includegraphics{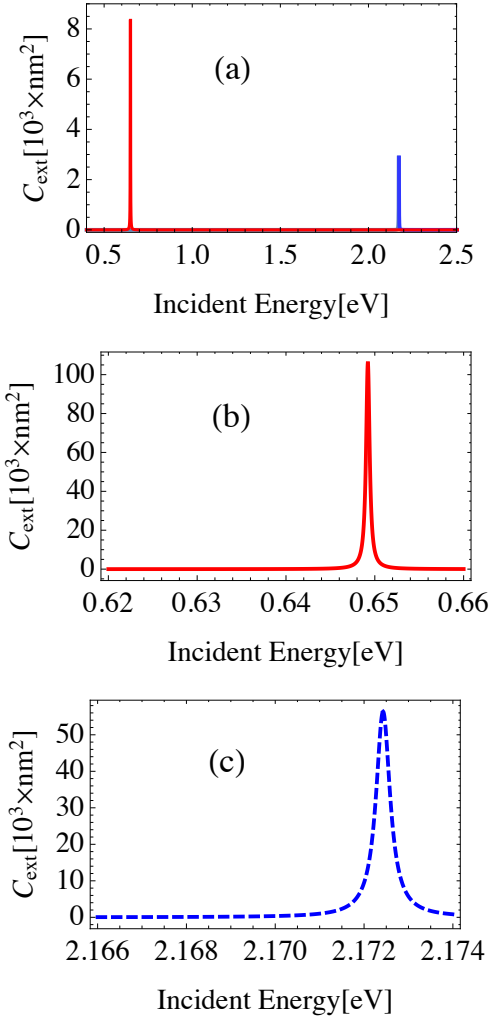}
\caption{\label{fig:2}The extinction cross section of GND of $L_{z}=7nm$ and $L_{x}=0.5nm$ with Fermi energy of $1.36 eV$, temperature of $300 K $and carrier's mobility of $10^{4}cm^{-2}/Vs$ embedded in GaAs, induced by z-polarized light (red-solid) and  x-polarized light (blue-dashed). Panels (b) and (c) show the width of the two resonances in (a).}
\end{figure}
 \section{Numerical Simulation and Discussion}
 \label{sec3}
We consider GND with radius of $L_{z}=7nm$ and thickness of $L_{x}=0.5nm$ to be able to neglect edge effects \cite{christensen2014classical}. The conductivity of GND is calculated at Fermi energy of $1.36 eV$, temperature of $300 K$and carrier's mobility of $10^{4}cm^{-2}/Vs$ which is equivalent to plasma energy of 7eV. For these parameters of GND embedded in GaAs background of $\epsilon_{b}=12.9$, the LSPRs are $\hbar\omega_{sp}^z=0.6481 eV$ and $\hbar\omega_{sp}^x=2.1724 eV$ as shown in Fig.~\ref{fig:2}. We also consider a spherical silver MNP of radius $R_{M}=15nm$, $\epsilon_{\infty}=5.7$, $\omega_{p}=1.36\times 10^{16} s^{-1}$ and $\gamma_{M}=10^{14}s^{-1}$. Note that at this value of MNP size, the polarizability is comparable to that of GND. The center-to-center distances are determined in terms of $R_{GM}$ and $\theta_{Q}=\phi_{1}+\phi_{2}$ as follows:
\begin{subequations}
\label{eq:eq11}
\begin{eqnarray}
R_{QG}=\left(\frac{sin\theta_{M}}{sin\theta_{Q}}\right) R_{GM},\label{11a}
\end{eqnarray}
\begin{equation}
R_{QM}=\left(\frac{sin\theta_{G}}{sin\theta_{Q}}\right) R_{GM}\label{11b}
\end{equation}
\end{subequations}
\begin{figure}
\includegraphics{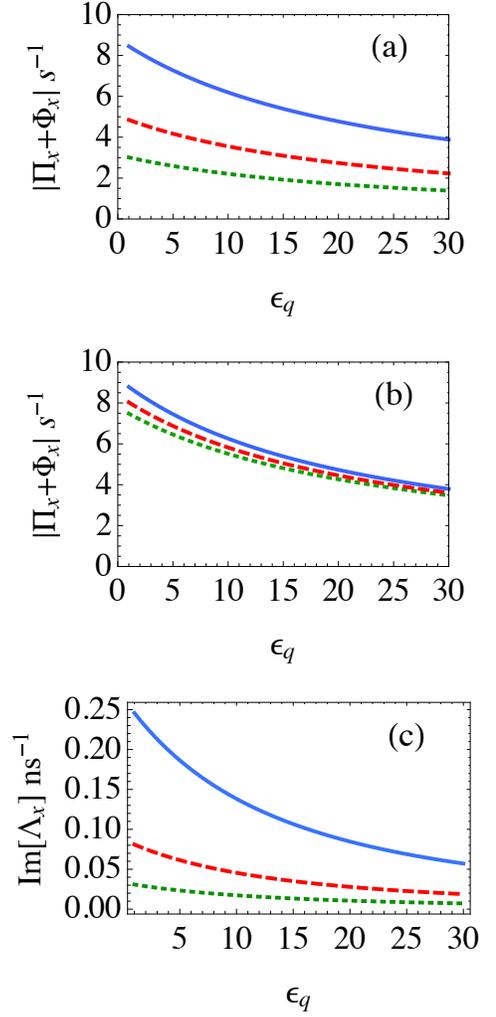}
\caption{\label{fig:3}The dependence of the enhancement factor of Rabi frequency (a,b) in the proposed MNP-GND-QD hybrid system induced by probe field at $2.1724 eV$, on the QD dielectric constant, (a): $\theta_{M}=0.26rad$ (solid), $\theta_{M}=0.3rad$ (dashed) and $\theta_{M}=0.36rad$ (dotted) with $R_{GM}=30nm$, $\theta_{G}=1rad$. (b): $R_{GM}=22nm-\theta_{M}=0.348rad$ (solid), $R_{GM}=26nm-\theta_{M}=0.315rad$ (dashed) and $R_{GM}=33nm-\theta_{M}=0.243rad$ (dotted) with $R_{QG}=8nm$ and $\theta_{G}=1rad$. (c): The dipole term of decay rate corresponding to the cases presented in (a).}
\end{figure}

Let's now examine the best QD material that can support the enhancement factor of probe field Rabi frequency i.e., $|\Pi_{x}+\Phi_{x}|$ and the dipole term of the decay rate, $Im\left[\Lambda_{x}(\rho_{22}-\rho_{11})\right]$. Fig.~\ref{fig:3}a and c show the dependence of these quantities on the dielectric constant of QD for incident photon of energy $ 2.1724eV$ with different values of $\theta_{M}$ at constant $\theta_{G}$ and $R_{GM}$. Note that, decreasing $\theta_{M}$ leads to decreasing $R_{QG}$ and increasing $R_{QM}$. It can be seen that for relatively small $\theta_{M}$, the Rabi frequency of probe field and the decay rate are enhanced due to the DDI. To show the affect of MNP on this enhancement, we keep $R_{QG}$ constant at 8nm and examine the enhancement factor of probe field Rabi frequency at different values of $R_{GM}$ by adjusting the angle $\theta_{M}$ as shown in Fig.~\ref{fig:3}b. It is remarkable that, as MNP comes closer to GND, the Rabi frequency of probe field is enhanced significantly for small $\epsilon_{q}$. Therefore, we choose  CdSe/GaAs self-assembled QDs since it  has  a resonance emission band in $2.165 eV$ \cite{mackowski2004optical} as well as relatively small dielectric constant for CdSe i.e., $\epsilon_{q}=6.5$. The transition energies in the QD are set as $\hbar \omega_{12}=2.172434 eV$ and $\hbar \omega_{13}=2.172432 eV$, to be nearly degenerate resulting in small two-photon detuning and thus enhanced energy transfer efficiency \cite{bederson2001advances}, and lie near to that of surface plasmon polaritons of GND. The decay widths of the QD  are taken to be $\Gamma_{12}=\Gamma_{13}=1.3\mu eV$ and $\Gamma_{32}=0.3\Gamma_{12}$ \cite{wang2017modification}. Note that the control field does not couple with surface plasmons since it is applied along the  z direction and $\omega_{13}$ is far detuned from $\omega_{sp}^{z}$.  

\begin{figure*}
\includegraphics{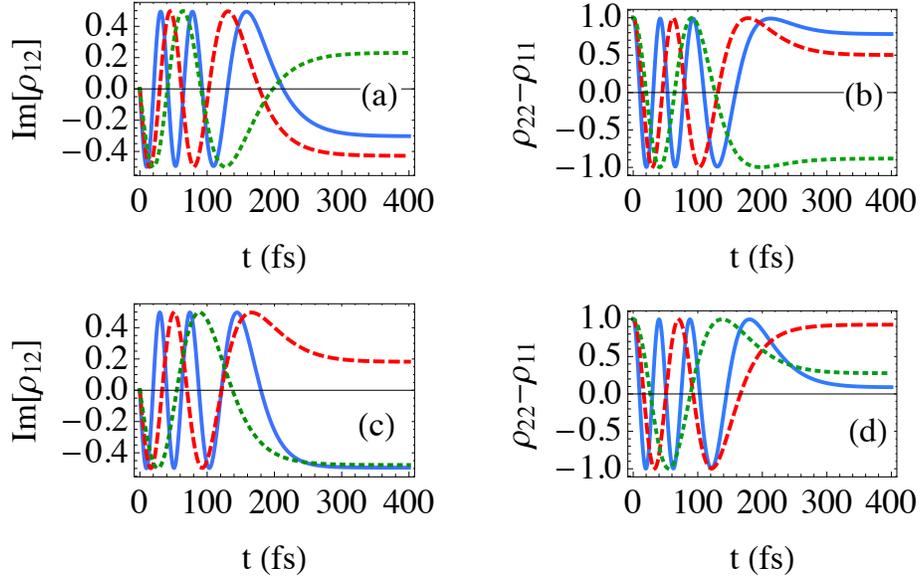}
\caption{\label{fig:4}The CdSe/GaAS self-assembled QDs decay rate (a,c) and population inversion (b,d) in the proposed MNP-GND-QD hybrid system for;(a,b): $R_{GM}=26nm$ (solid), $R_{GM}=29nm$ (dashed) and $R_{GM}=32nm$ (dotted) with $\theta_{M}=0.3 rad$ and $\theta_{G}=1rad $,.(c,d): $\theta_{M}=0.26 rad$ (solid), $\theta_{M}=0.3rad$ (dashed) and $\theta_{M}=0.36 rad$  (dotted) with $ R_{GM}=30nm$ and $\theta_{G}=1rad$ excited by probe and control pulses of $\Delta\tau_{p}=100 fs$ and  $\Delta\tau_{c}=30 fs$, $t_{p}=1fs$, $\delta=\Delta\tau_{c}$, and detuning of $\Gamma_{p}=1.3\mu eV$ and $\Gamma_{c}=0$}
\end{figure*}

 Let's now investigate the energy transfer between excitons and surface plasmon polaritons via the probe field. Note that the control field is applied to monitor the energy transfer. The two fields are chosen as Gaussian pulses of the form \cite{carreno2014plasmon}:  
\begin{subequations}
\label{eq:12}
\begin{eqnarray}
\Omega_{p}(t)=\frac{\Omega_{0}}{\sqrt{2\pi \Delta\tau_{p}^{2}}}e^{-(t-t_{p})^2/2\Delta\tau_{p}^{2}},\label{12a}
\end{eqnarray}
\begin{equation}
\Omega_{c}(t)=\frac{\Omega_{0}}{\sqrt{2\pi \Delta\tau_{c}^{2}}}e^{-(t-t_{p}-\delta)^2/2\Delta\tau_{c}^{2}}\label{12b}
\end{equation}
\end{subequations}
where $\Omega_{0}$ is the normalized peak amplitude that measures the pulse area, $\Delta\tau_{p}(\Delta\tau_{c})$is the FWHM time duration of the probe pulse (control pulse), $t_{p}$ is the center of the pulse and $\delta$ is the time delay between the two pulses. Suppose $\Delta\tau_{p}=100 fs$, $\Delta\tau_{c}=30 fs$ and the center of the pulse is taken to be $t_{p}=1fs$. The probe pulse is applied just as the control pulse is switched off, that is $\delta=\Delta\tau_{c}$. 
FIG.~\ref{fig:4} depicts the decay rate and population inversion in the QD induced via coupling excitonic transition $\ket{1}\leftrightarrow \ket{2}$ with the surface plasmons $\hbar \omega_{sp}^{x}$, obtained by solving for the time evolution of the density matrix elements that describes the energy transfer within the system. We check to what extent we can control the decay rate and amplification in the system by its geometrical conditions. We present our main results for the effect of $\theta_{M}$ and $R_{GM}$ on the energy transfer between excitons and plasmons in Fig.~\ref{fig:4}. It can be seen that at small distances between MNP and GND,  the decay rate oscillates rapidly associated with transient population inversion over few tens of femtoseconds. As $\theta_{M}$  decreases, the decay rate is enhanced leading to almost vanished steady state population inversion as shown in FIG.~\ref{fig:4}c, d. This is because small values of center-to-center distances $R_{QG}$ are associated with small $\theta_{M}$. However, the system can demonstrate strong energy transfer by tailoring $\theta_{M}$ and $R_{GM}$ resulting in a significant amplification for surface plasmons. 

\begin{figure}
\includegraphics{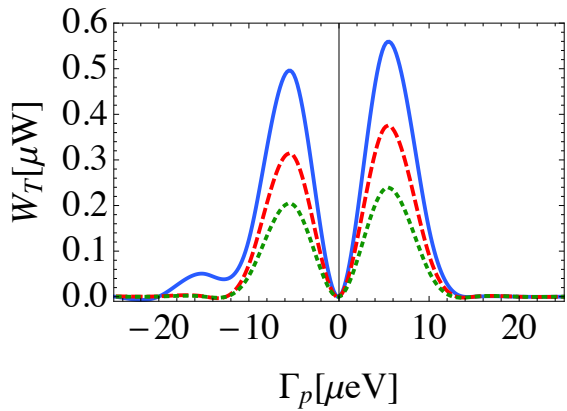}
\caption{\label{fig:5} : The energy transfer rate between the QD and the GND in the proposed MNP-GND-QD hybrid system with $|\Pi_{x}+\Phi_{x}|=6$ (solid), $|\Pi_{x}+\Phi_{x}|=5.21$ (dashed) and $|\Pi_{x}+\Phi_{x}|=4.37$ (dotted), excited by probe and control pulses of intensity $0.2$ and $2.5$ $TW cm^{-2}$respectively with $\Gamma_{c}=0$.}
\end{figure}

To highlight the enhancement in the energy transfer rate within the proposed MNP-GND-QD hybrid system, let's now calculate the energy transfer rate between the QD and the GND in the presence of MNP. The energy transfer rate is given as the volume integral of the scalar product of the electric field felt by the GND and its current density. By taking the time average of this rate over the period of oscillation we get \cite{singh2016excitonic,sarid2010modern}:
\begin{widetext}
\begin{equation}
W_{T}=-2\times 10^{-4}\epsilon_{0} \frac{\omega_{12}\hbar^{2}}{2\pi \mu_{12}^{2}} Re[\alpha_{G}^{x}]{Re\left[\frac{\Omega_{0}}{\sqrt{2\pi\Delta\tau_{p}^{2}}}+\frac{(3 cos\theta_{Q}/2-1)\rho_{12}\mu_{12}^{2}}{4\pi \epsilon_{0} \hbar\epsilon^{*}\epsilon_{b} R^3}-\frac{\alpha_{M}}{4\pi \epsilon^{*} R_{MG}^3}\frac{\Omega_{0}}{\sqrt{2\pi\Delta\tau_{p}^{2}}}\right]}^{2}
 \label{eq:13}
\end{equation}
\end{widetext} 
where the three terms represent the total field felt by the GND due to the probe field, the QD and the MNP respectively. We first solve numerically the equations of motion for the density matrix elements at steady state for different values of the probe field detuning,$\Gamma_{p}$. We consider that the intensity of probe and control fields are $0.2$ and $2.5$ $TW cm^{-2}$ respectively. The rate of energy exchange between plasmons and excitons is illustrated in Fig.~\ref{fig:5} at different values of the enhancement factor of probe field Rabi frequency. As we expect, relatively large values of energy exchange rate are obtained for the cases of large enhancement factor, $|\Pi_{x}+\Phi_{x}|$ and thus, small $R_{QG}$. The effect of increasing the energy transfer by decreasing the distances between QD and graphene has been observed experimentally by several research groups with different types of QDs \cite{dong2010fluorescence,chen2010energy}. Moreover, strong fluorescence quenching was observed for QDs deposited on the graphene sheet due to the large energy transfer rate between QD and graphene \cite{chen2010energy}. A double-peaked dependence on the probe field detuning can be explained using the theory of dressed states arising from the strong coupling when the QD is close to GND \cite {ficek2005quantum,cox2012dipole}, since the values of enhancement factor used in Fig.~\ref{fig:5} are associated with small $R_{QG}$. We observe that the difference between the resonance of positive and negative probe field detuning increases as $|\Pi_{x}+\Phi_{x}|$ increases due to the relatively large decay rate, enhanced by the DDI as predicted in Fig.~\ref{fig:3}a, c.  

\begin{figure}
\includegraphics{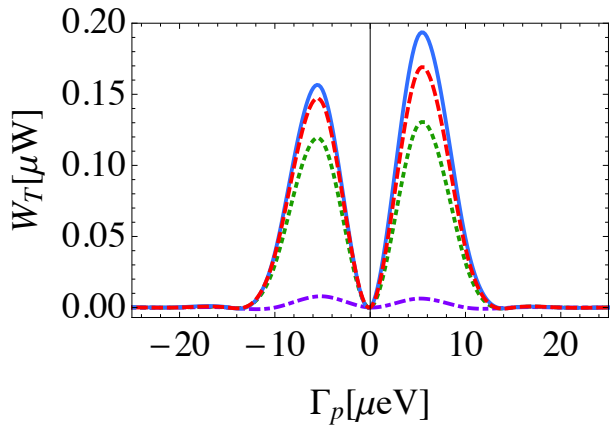}
\caption{\label{fig:6} : The energy transfer rate between the QD and the GND in the proposed MNP-GND-QD hybrid system with $R_{GM}=25nm$, $\theta_{M}=0.32rad$, $\theta_{G}=1rad$, $R_{M}=18nm$ (solid), $R_{M}=15nm$ (dashed), $R_{M}=13nm$ (dotted) and $R_{M}=7nm$ (dot-dashed), excited by probe and control pulses of intensity $0.2$ and $2.5$ $TW cm^{-2}$respectively with $\Gamma_{c}=0$.}
\end{figure}

To underline the role of MNP in the energy exchange between QD and GND, we investigate the power transfer with different small sizes of MNPs, in which the near-field approximation is  valid, as shown in Fig.~\ref{fig:6}. The center-to-center distances between GND and MNP are kept at $25nm$ with $\theta_{M}=0.3rad$ and $\theta_{G}=1rad$. It is clear that the power transfer significantly depends on the size of MNP. As the size of MNP increases, the rate of energy exchange between plasmons and excitons increases due to the relatively large corresponding enhancement factor. It is remarkable that the resonance of positive probe field detuning is enhanced at the expense of that corresponding to negative one as the size of MNP increases. This can be interpreted by the enhanced decay rate for large size of MNPs.

Compared with the femtowatt energy exchange rates that have been shown for the hybrid system consisting of QD and GND of 6.02 eV plasma energy in the infrared regime with pumping intensity of  $31GW cm^{-2}$ \cite{cox2012dipole}, it is clear that the presence of MNP of size exhibiting polarizability comparable to that of a highly doped GND, e.g.  has plasma energy of 7 eV as in our case, with self-assembled QD of small dielectric constant interacting with ultrashort probe and control pulses can lead to a pronounced enhancement in the energy exchange rate between GND and QD. Our results are supported by those found experimentally by Alonso-Gonzalez et al that the plasmons of graphene can be launched and controlled effectively with resonant metal antennas \cite{alonso2014controlling}. 

\section{Conclusions}
\label{sec4}
We have investigated the DDI and energy transfer in a MNP-GND-QD hybrid system in the optical region with CdSe/GaAs self- assembled QDs modeled as three-level $\Lambda$ configuration atoms excited by two ultrashort pulses.  Our simulation predicts that the Rabi frequency of the probe field and the decay rate are enhanced by the DDI between the components of the hybrid system and can be controlled by the geometrical characteristics of the system as well as the dielectric environment. We have demonstrated ultrafast decay rate and transient population inversion of the system over few tens of femtoseconds and can be adjusted by the inclination angles of MNP and GND with respect to QD and the center-to-center distances between the components of the system. We have obtained a large energy exchange rate between excitons and plasmons within the proposed MNP-GND-QD hybrid system for the cases of relatively large enhancement factor of probe field Rabi frequency demonstrating a double-peaked dependence on the probe detuning. Moreover, the energy exchange rate  significantly depends on the size of MNPs. A relatively large size of MNPs, within the limit of the near-field approximation,  that exhibits large enhancement factor of probe field Rabi frequency is preferable to get large energy exchange rate between excitons and plasmons within the MNP-GND-QD hybrid system. 

Our numerical simulation results, based on the DDI interactions between the components of our proposed hybrid system, provide motivation for future experimental and theoretical investigations on the hybrid nanocomposites due to the controllable ultrafast dynamics that can be demonstrated by these systems. Moreover, our results could prove useful in the development of controllable nanosensors, energy transfer and energy storage devices in the nanoscale and all-optical nanoswitches. The present results could be employed to build efficient and controllable plasmonic amplifiers.
 
\begin{acknowledgments}
This Project was supported by King Saud University, Deanship of Scientific Research, Research Chairs.
\end{acknowledgments}
\nocite{*}

\bibliography{apssamp}

\end{document}